# Metamorphic Proteins in light of Anfinsen's Dogma

Jorge A. Vila

IMASL-CONICET, Universidad Nacional de San Luis, Ejército de Los Andes 950, 5700 San Luis, Argentina.

**Corresponding author**
Jorge A. Vila jv84@cornell.edu
https://orcid.org/0000-0001-7557-9350

**ABSTRACT**

It is a common belief that metamorphic proteins challenge Anfinsen's thermodynamic hypothesis (or dogma). Here we argue against this view aims to show that metamorphic proteins not just fulfill Anfinsen's dogma but also exhibit marginal stability comparable to that seen on biomolecules and macromolecular complexes. This work contributes to our general understanding of protein classification and may spur significant progress in our effort to analyze protein evolvability.

The term "metamorphic proteins" was coined by Murzin[1] after observe that some proteins "…*adopt different folded conformations for the same amino acid sequence in native conditions*…" The first and perhaps most outstanding example of metamorphic protein was discussed by Murzin[1] itself who noted that "… *the chemokine lymphotactin (Ltn) studied by Tuinstra et al. adopts two distinct folds at equilibrium in physiological conditions, and interconversion between the* *conformers involves almost complete restructuring of its hydrogen bond network and other stabilizing interactions…*"

Later on, a plethora of papers analyze and discuss a growing number of, the so-called, metamorphic proteins.[2] Although it is not in the spirit of this Letter to review all the existing literature about this topic we will discuss some cases of metamorphic proteins in light of the Anfinsen's dogma[3] and the aftereffect, namely, the existence of an upper bound limit to the protein marginal stability.

As we will illustrate below, with a few examples, the experimental condition (*milieu*) plays a central role in the analysis of metamorphic proteins. In this regard, we identify two possible scenarios of protein folding:

*Scenario 1*.- For a given experimental milieu the validity of the Anfinsen thermodynamic hypothesis requires that the Gibbs functional $G(\mathcal{Z}\{\phi,\psi,\chi\})_{mileiu}$ has the lowest free-energy minimum with respect to *all* possible distributions of $\mathcal{Z}\{\phi,\psi,\chi\}$, where $(\phi,\psi,\chi)$ are the protein torsional angles. Around this lowest free-energy minimum, *viz.*, the native-state, there is an ensemble of native-folds that coexist in fast dynamics equilibrium.[4,5] Regarding the coexistence between native-folds we have been able to prove the existence of an upper limit to the marginal stability of monomeric globular proteins, namely $\approx 7.4$ Kcal/mol.[6] This free-energy restraint applies to any fold-class, sequence, or protein size, while Anfinsen's dogma is fulfilled.[7]

*Scenario 2*.- Given two, or more, experimental milieus*,* then the analysis for the *scenario 1* (see above) apply for each of them. Proteins that fold under this scenario may lead to structural metamorphosis. Indeed, this happens when there are changes in some critical parameters of the *milieu*. Let us illustrate it with two examples. The intracellular chloride ion channel protein (CLIC1) shows a metamorphic conformational shift in response to a change in the local environment, i.e., driven by a change in the oxidizing conditions;[8,9] similarly, the *E. coli* elongation factor RfaH shows a significant structural change upon ligand binding to the nontemplate DNA.[9,10] Both proteins are considered as metamorphic proteins. Undoubtedly, there is no challenge to Anfinsen's dogma here since a native-state exist for a given, fixed, mileiu.

On one hand, proteins that fold under scenario 1 may lead to the coexistence between "greatly-similar" native-folds, *e.g.*, for ubiquitin[5] as well as for many other monomeric globular proteins, such as α-chymotrypsin, ribonuclease A, cytochrome c, etc.[11] For any protein of this class, we have proved the existence of an upper bound to protein's marginal stability.[6] On the other hand, it may also happen the coexistence between "highly-dissimilar" native-folds (as for metamorphic proteins), *e.g.*, protein A,[12] the spindle checkpoint protein Mad-2,[13] or the human chemokine lymphotactin (Ltn).[14] Then, the following questions arise: do any of these metamorphic proteins challenge Anfinsen's dogma? if they do not, are these metamorphic proteins marginally stable? Let us briefly analyze each of them.

1. The 3D structure of the B-domain of protein A (from Staphylococcus aureus) has been used as a protein folding model and solved at the atomic level (determining the lowest free-energy minimum of the Gibbs functional) by using different force-fields.[12, 15-18] All these theoretical studies show the coexistence between the native-state of protein A with its mirror-image, an energetically competitive conformation;[15-18] indeed, the coexistence occurs within a narrow range of free energy variation, namely $-4.2 < \Delta G < -1.5$ Kcal/mol.

2. Protein Mad2, at 37°C, spontaneously converts between two native-folds with slow kinetics, e.g., at equilibrium, the $\Delta G$ of the reaction is about 1.4 Kcal/mol[13,19] indicating that one of them is the lowest energy state, hence, the native state.

3. Protein Ltn, under a similar experimental condition that for protein Mad2, namely 37ºC and 150 mM NaCl, shows a rapid interconversion between two equally populated conformers, indicating that the coexistent native-folds have about the same free-energy.[13] Additionally, as noted by Luo & Yu,[13] the equilibrium between equally populated conformers (for both Ltn and Mad2) reveals a thermodynamic-controlled process; hence, assuring that Anfinsen's dogma rules the folding of these proteins.

For all three cases, the Anfinsen's dogma is accomplished because it does not refer to a unique structure but to an ensemble of structures around the lowest free-energy minimum of the Gibbs functional, namely the native-state. As a result, the metamorphic protein's marginal stability, in all cases, is consistent with the existence of an upper bound of around 7.4 Kcal/mol.[6]

Overall, we are in a condition to conclude, without a doubt, that there are no mysteries, no anomalies, no paradigm behind the existence of metamorphic proteins. They simply fulfill the Anfinsen's thermodynamic hypothesis and, hence, exhibit marginal stability comparable to that seen on biomolecules and macromolecular complexes.[7]

Lastly, it is worth warning that prion proteins and intrinsically disordered proteins fail to form a stable 3D structure and, hence, they do challenge Anfinsen's dogma and, hence, the existence of an upper bound protein marginal stability.

**Notes**

The author declares no competing financial interest.

**ACKNOWLEDGMENT**

The author acknowledges financial support from the IMASL-CONICET (PIP-0087) and ANPCyT (PICT-0767; PICT-2212), Argentina.

**TOC GRAPHICS**

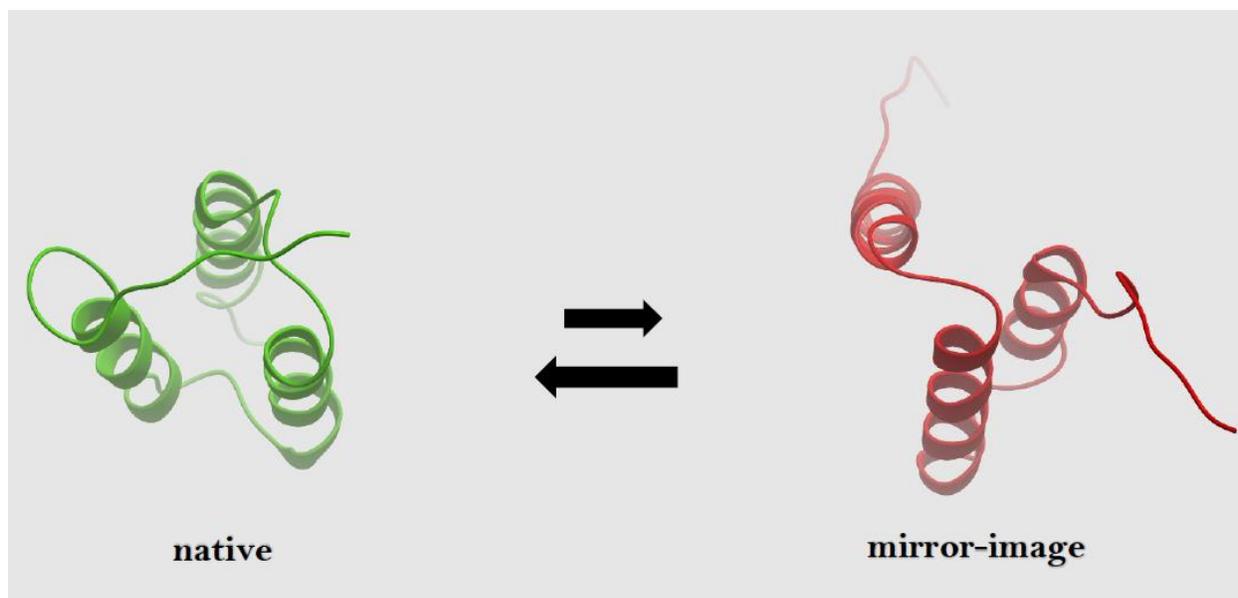